# The need for focused, hard X-ray investigations of the Sun

Lead author: Lindsay Glesener, University of Minnesota
Full author list: See spreadsheet

Understanding the nature of energetic particles in the solar atmosphere is one of the most important outstanding problems in heliophysics. Flare-accelerated particles compose a huge fraction of the flare energy budget; they have large influences on how events develop; they are an important source of high-energy particles found in the heliosphere; and they are the single most important corollary to other areas of high-energy astrophysics. Despite the importance of this area of study, this topic has in the past decade received only a small fraction of the resources necessary for a full investigation. For example, **NASA has selected no new Explorer-class instrument in the past *two decades* that is capable of examining this topic.** The advances that are currently being made in understanding flare-accelerated electrons are largely undertaken with data from EOVSA (NSF), STIX (ESA), and NuSTAR (NASA Astrophysics). This is despite the inclusion in the previous Heliophysics decadal survey of the FOXSI concept as part of the SEE2020 mission, and also despite NASA's having invested heavily in readying the *technology* for such an instrument via four flights of the FOXSI sounding rocket experiment. Due to that investment, the instrumentation stands ready to implement a hard X-ray mission to investigate flare-accelerated electrons. This white paper describes the scientific motivation for why this venture should be undertaken soon.

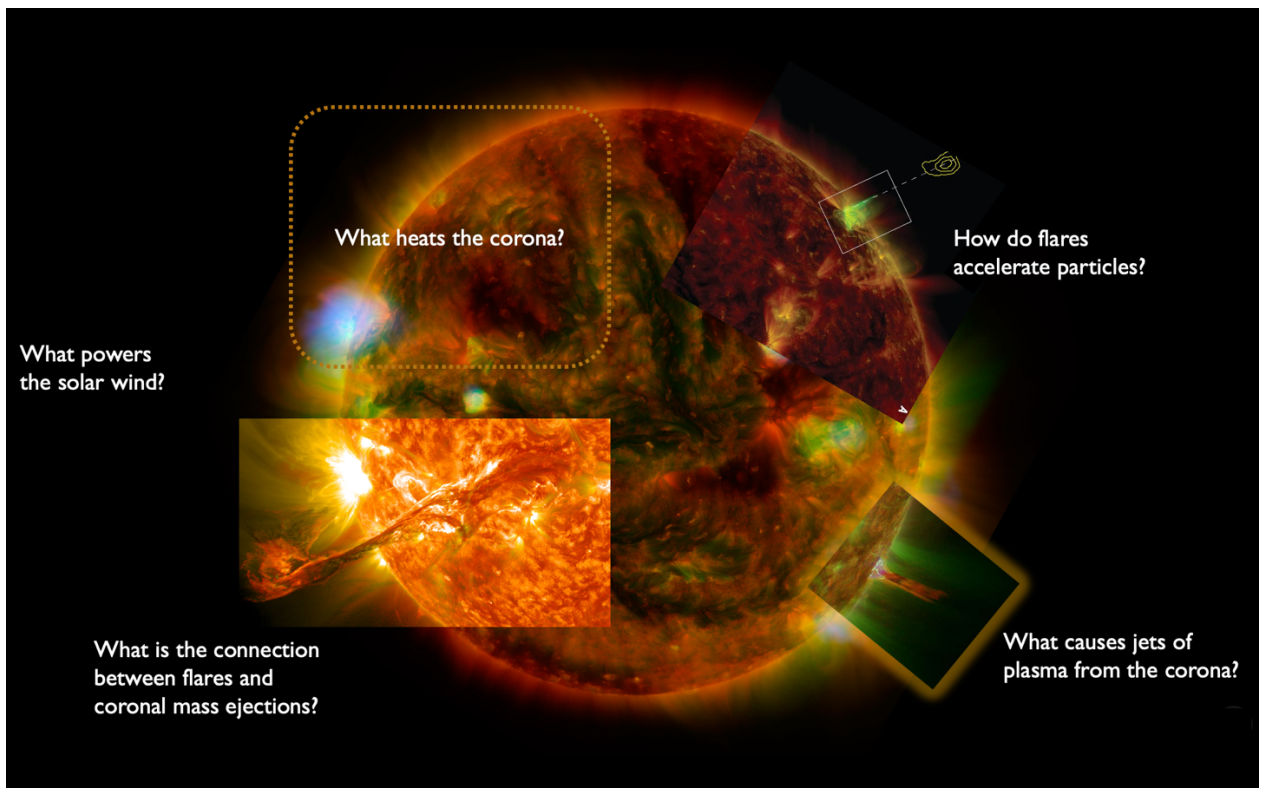

*Image is a compilation of works by NASA/JPL-Caltech/GSFC/JAXA, SDO/Wiessinger, SDO/SDTC/Wyper, and Science/Chen.*

# The need to understand electron acceleration at the Sun

**The problem of particle acceleration in a magnetized plasma is a ubiquitous issue throughout astrophysics.** Understanding how high-energy particles can be produced is fundamental to understanding, for example, the nature of supernova remnants; magnetar atmospheres, black hole jets, and the sources of cosmic rays, as well as how flares accelerate particles on the Sun. Some acceleration processes likely operate in many of these regimes, for example shock acceleration, various forms of Fermi acceleration, and direct acceleration by reconnection electric fields (e.g. Drake et al. 2006, Petrosian 2012, Kong et al. 2019, Li et al. 2021). Of these astrophysical contexts, the Sun is the one where the particle acceleration environments and mechanisms can be most fully probed. This includes observations of the accelerated particles themselves, via remote-sensing X-rays or microwaves, or in situ detection of the particles that escape into the heliosphere. It also includes a multiwavelength, high resolution view of the plasma environment in which the particle acceleration occurs, specifically volumes, temperatures, densities, and sometimes magnetic fields. (These could be probed even more by multi-viewpoint observations, which are conceivable for the Sun; see, for example, white papers on the *COMPLETE* concept, Caspi et al. 2022.) For more on the connections between high-energy solar physics and astrophysics, see the white paper by Vievering et al. (2022).

**Similarly, particle acceleration links together many realms of heliophysics.** Questions about how particles are accelerated start at the Sun, but they don't end there. Particles are energized in the solar wind, at interplanetary shocks and bow shocks, and at diverse locations within planetary magnetospheres. X-ray, microwave, and in situ observations of some of these contexts exist (e.g. Millan et al. 2013; Breneman et al. 2017), and thus remote sensing of the Sun is the best way to unify this science across a diverse set of environments. What's more, particles accelerated at the Sun are found across the heliosphere. This offers a unique opportunity to study high-energy particles across a wide range of their lifetimes and across interplanetary-scale ranges. In the case of accelerated electrons, study could begin at the Sun with remote-sensing (X-ray and microwave) measurements of acceleration sites; this is followed by radio observations of Type III emission from accelerated electrons leaving the Sun; and finally, electrons can be detected in-situ via particle detectors or by the plasma waves they create as they traverse through the heliosphere (see **Figure 1**). Such study would provide a complete picture of how electrons are accelerated at the Sun and all the ways in which they evolve as they propagate into interplanetary space. *This is especially time critical as Parker Solar Probe and Solar Orbiter provide new insight on the nature of energetic particles and the magnetic topology of the inner heliosphere* (e.g. Bale et al. 2022, Phan et al. 2022). Direct HXR studies would provide quantitative estimates of particles leaving the Sun. But these studies are not possible with existing instruments; hard X-ray instruments that operated in the past or present have not had the dynamic range necessary to measure acceleration locations, nor to measure emission from escaping beams (e.g. Saint-Hilaire et al. 2009).

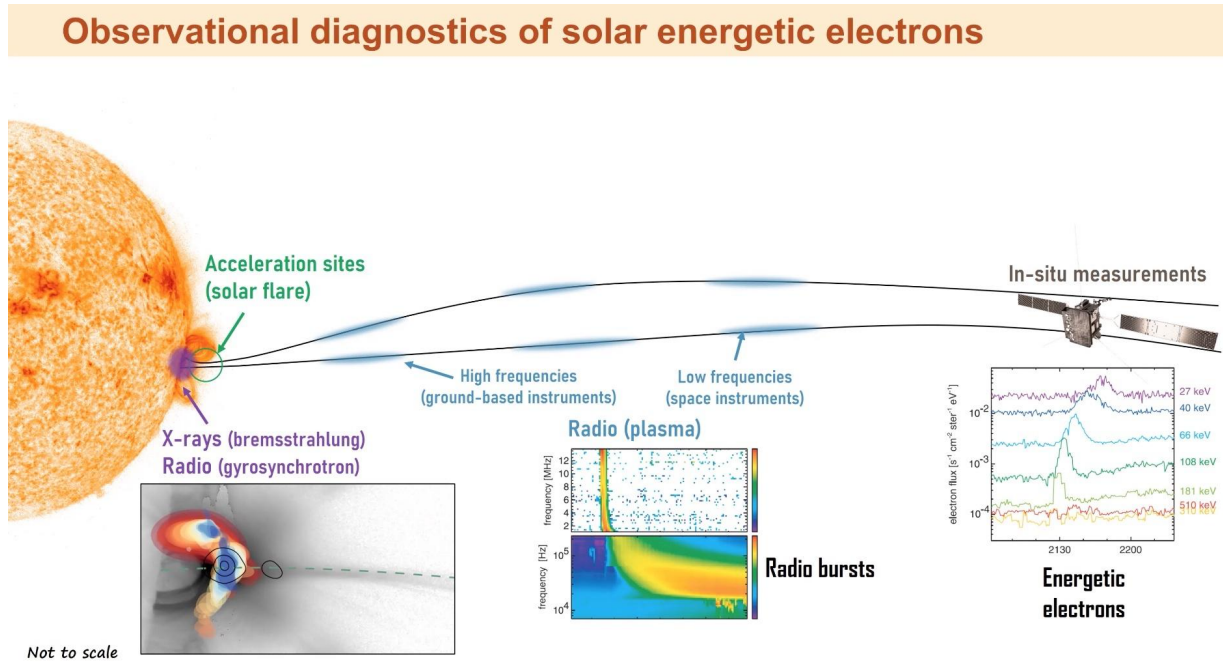

**Figure 1:** HXR observation of electrons as they are accelerated at flare sites is an important piece of studying the evolution of these particles throughout the heliosphere. *Figure from S. Musset.*

Accelerated electrons account for a significant fraction of the total energy budget of a flare and/or eruptive event. Emslie et al. (2012) examined 38 eruptive flares and found that the flare-accelerated electrons and ions are sufficiently energetic to supply the bolometric energy radiated across all wavelengths throughout the event. This is strong evidence for the thick target solar flare model put forth by Brown (1971), in which accelerated particles are the primary means by which energy is transferred from coronal magnetic reconnection sites to the flare footpoints, and thus the primary means by which solar flares are heated. In this way, accelerated electrons (and ions) are uniquely important to study because they influence everything that comes afterward: the heating of the flare, the response of the lower solar atmosphere to the release of magnetic energy, the heating of any associated coronal mass ejection (CME), and the emergence of coronal jets. The lower solar atmosphere is where the bulk of the energy from accelerated electrons gets deposited, and so it is critical to thoroughly study the coupling between accelerated particles, their propagation, and the atmospheric response (e.g., white paper by Allred et al. 2022) and also the energy budget of the flare during the impulsive phase (e.g. white paper by Kerr et al. 2022).

## Limitations of past and current hard X-ray instruments for solar observation

Since accelerated electrons emit bremsstrahlung X-rays cleanly observable in the hard X-ray (HXR) range, HXRs are one of the few tools available to probe these populations. (Microwave observations constitute another critical tool; microwaves and HXRs are highly complementary due to their dependence on magnetic field and densities, respectively, e.g. White et al., 2011. Also see decadal white papers by Gary et al. 2022 and Chen et al. 2022.) Unfortunately, most aspects discussed in the previous section are difficult to study with the limited imaging dynamic range available with previous (indirect) solar-dedicated HXR imagers. Since the decommissioning of the *Reuven Ramaty High Energy Solar Spectroscopic Imager* (*RHESSI*; Lin et al., 2002) in 2018, there is no solar-dedicated HXR observatory. *RHESSI* made unparalleled advances in understanding particle acceleration in flares but, given its indirect imaging method (Hurford et al., 2003), lacked the necessary sensitivity to study faint coronal particle acceleration sites in all but the brightest flares, and lacked the imaging dynamic range to regularly study coronal sources in the presence of bright footpoints (Krucker et al., 2014).

There is evidence (via occulted flares) that all or almost all flares have coronal nonthermal sources (Krucker & Lin 2008). Coronal nonthermal sources are usually 10-100 times fainter than flare footpoints in HXRs (see Fig. 6.1 of Glesener 2012), and this is beyond *RHESSI's* typical imaging dynamic range. For example, **Figure 2** shows examples of faint coronal HXR in a coronal mass ejection (CME) and a coronal jet observed by *RHESSI*. In both cases accelerated electrons were found to play a large role in the energetics of the events, but both cases are rare HXR observations due to *RHESSI's* limited sensitivity and dynamic range. In fact, neither of these observations was possible without partial occultation of the flare by the solar limb. **Systematic study of these coronal HXR sources is critical in order to link accelerated particles in the low corona to those in the heliosphere, and to understand how flares accelerate particles with extraordinary efficiency.**

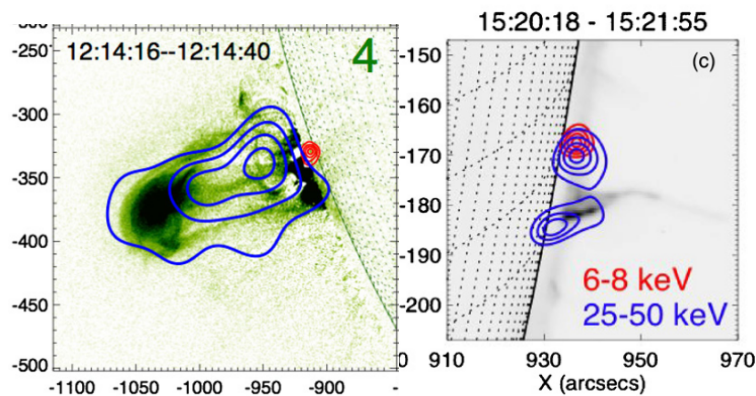

**Figure 2:** Rare examples of nonthermal HXRs observed in *RHESSI* eruptive events. (Left) Accelerated electrons were found filling the core of a CME shortly after its initiation and were found to be responsible for the heating of that core. Figure from Glesener et al. (2013). These faint HXRs were only visible to *RHESSI* via partial occultation and a nonstandard imaging algorithm (called two-step CLEAN). (Right) Accelerated electrons were found to be co-spatial with an emerging coronal jet, showing that electrons accelerated at the flare site have access to the jet path and can emerge into interplanetary space. Again, the event needed to be partly occulted for *RHESSI* to measure this. Figure from Glesener et al. (2012).

The *Fermi* Gamma-ray Telescope often observes HXR flares but produces only spectra, no imaging, and often experiences severe pileup during bright flares since it is not optimized for the Sun. Similar issues apply to the Konus instrument aboard the *WIND* spacecraft. The Large Area Telescope aboard *Fermi* has provided remarkable insight into the extremely high energy tails of flare-accelerated particle distributions (e.g. Pesce-Rollins et al. 2015, Ajello et al. 2021) but cannot measure the parts of the distributions containing most of the energy (and is again not solar-optimized). The only currently operating solar-dedicated HXR instrument is the Spectrometer Telescope for Imaging X-rays (STIX; Krucker et al., 2020) onboard ESA's *Solar Orbiter* mission.

The next expected mission is the Hard X-ray Imager (HXI) onboard the *Advanced Space-based Solar Observatory (ASO-S*, launch 2022; Zhang et al. 2019). While STIX is already producing novel science on flare electron distributions (e.g. Battaglia et al. 2021, Massa et al. 2022), both STIX and HXI are indirect imagers and will face similar problems in sensitivity and dynamic range as *RHESSI*, even at closer observing distances. Lastly, the *Aditya-L1* mission from the Indian Space Research Organisation will carry a HXR spectrometer, but without imaging.

**The limitations faced in the past using indirect imaging can now be overcome with modern instruments with direct HXR focusing**; photons in different locations of the detector can be analyzed separately in a straightforward way. The prime example of this method is the *Nuclear Spectroscopic Telescope Array (NuSTAR)*, a Small Explorer launched in 2012 and funded by NASA's Astrophysics (not Heliophysics) Division.

*NuSTAR* is 100 times more sensitive than its predecessors (Harrison et al. 2013). This sensitivity comes not only from the large effective area (800 cm$^2$) but much more so from the use of focusing mirrors to concentrate X-rays down to a small detector area. This results in a drastic reduction in background by several orders of magnitude as compared with non-focusing instruments. Grefenstette et al. (2016) compute a figure of merit for comparing performance between HXR telescopes consisting of a background to effective area ratio. This figure of merit is roughly a million times lower for *NuSTAR* than for *RHESSI*, meaning a flare giving the same count rate as the background is a million times fainter for *NuSTAR* than for *RHESSI*. This translates to a sensitivity increase of roughly 40-4000 for *NuSTAR* over *RHESSI*, depending on detector lifetime (aka the fraction of time for which the detector is available to detect photons).

Unfortunately, since *NuSTAR* is not optimized for the Sun, it experiences very low livetime in the presence of bright solar flux, meaning that only the faintest microflares and quiescent solar emission can be observed without saturating the detectors. Measurement of flares above approximately B class (i.e. microflares) is not possible with *NuSTAR*. Such low livetimes (often <2%) also significantly restrict the spectral dynamic range since solar spectra are usually quite steep. Furthermore, with *NuSTAR's* limited angular resolution (18 arcsec full width half max, 58 arcsec half power diameter), it does not have sufficient angular resolution to resolve most flares.

In summary, currently planned or operational solar-dedicated HXR missions use indirect imaging methods that are limited in sensitivity and imaging dynamic range. Note that none of these currently operating or planned missions are NASA Heliophysics missions. **The clear next step for understanding accelerated electrons at the Sun is with a direct-focusing HXR instrument.**

## Next step: Direct HXR imagers, optimized for the Sun

Due to prior technology development investment by NASA Heliophysics, the technology is already prepared for a direct-focusing solar HXR mission. The *Focusing Optics X-ray Solar Imager (FOXSI)* sounding rocket experiment *is* optimized for the Sun and has flown successfully 3 times (2012, 2014, and 2018; see Krucker et al. 2014, Glesener et al. 2016, Musset et al. 2019). As a sounding rocket experiment, observations are limited to ~6 minutes each, and it has not historically been possible to target a large flare due to the lack of capability to predict such flares. (In March 2024, *FOXSI-4* will attempt to target a big flare for the first time, along with the *Hi-C* and *SNIFS* experiments; see Buitrago-Casas et al. 2021.) With solar optimization, *FOXSI* achieves the full sensitivity available to direct imagers. Vievering et al. (2021) estimated that the *FOXSI-2* experiment measured a sub-A-class microflare (shown in **Figure 3**) with a sensitivity 5 times better than what *NuSTAR* could achieve for a similar flare (given *NuSTAR*'s livetime limitations), so that *FOXSI* improves even further upon the sensitivity mentioned for *NuSTAR* in the previous section. Figure 4 shows a brightness-hardness diagram of solar microflares from Duncan (2022), demonstrating that direct imaging can achieve ~5 orders of magnitude better sensitivity than *RHESSI* could for small microflares. A recent paper by Buitrago-Casas et al. (2022, accepted) calculates upper limits on quiet Sun HXR emission using *FOXSI* data. Achieved limits are slightly better than what was achievable with *RHESSI*, but the *FOXSI* limits come from only 6 minutes of solar observation, while *RHESSI* needed ~12 days to achieve a similar result! This stark contrast illustrates the tremendous sensitivity available to direct HXR imagers that are optimized for the unique conditions of solar observations.

Similarly, the utility of direct focusing HXR instruments for studying coronal heating has been demonstrated. Ishikawa et al. (2017) showed that incorporating *FOXSI* HXR data into a differential emission measure calculated using *Hinode*/XRT measurements significantly changed the estimate of the temperature distribution. Specifically, it was found that the inclusion of HXRs provided a much more stringent estimate on the amount of hot, flare-temperature plasma present in the distribution, but showed unequivocally that a small amount of that plasma is indeed present. This is an important conclusion that supports the theory of nanoflare heating of that active region. Marsh et al. (2018) went further by simulating various frequencies of heating of that same region, and found that in order to match the *FOXSI* HXR data, impulsive heating was indeed required.

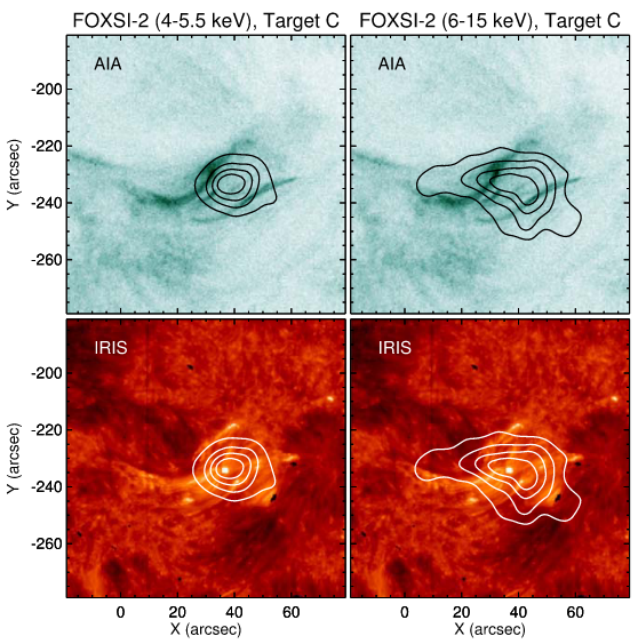

**Figure 3:** A solar microflare observed with *FOXSI-2*'s direct imaging method. Every photon in this image is measured individually, with the time of arrival, location, and energy recorded. In this way, scientists can produce lightcurves and spectra for any location within the field of view. The image shows *FOXSI-2* images overload on an AIA 94A image (top row) and an *IRIS* slit-jaw image (bottom row), in two different energy ranges. Figure from Vievering et al. (2021).

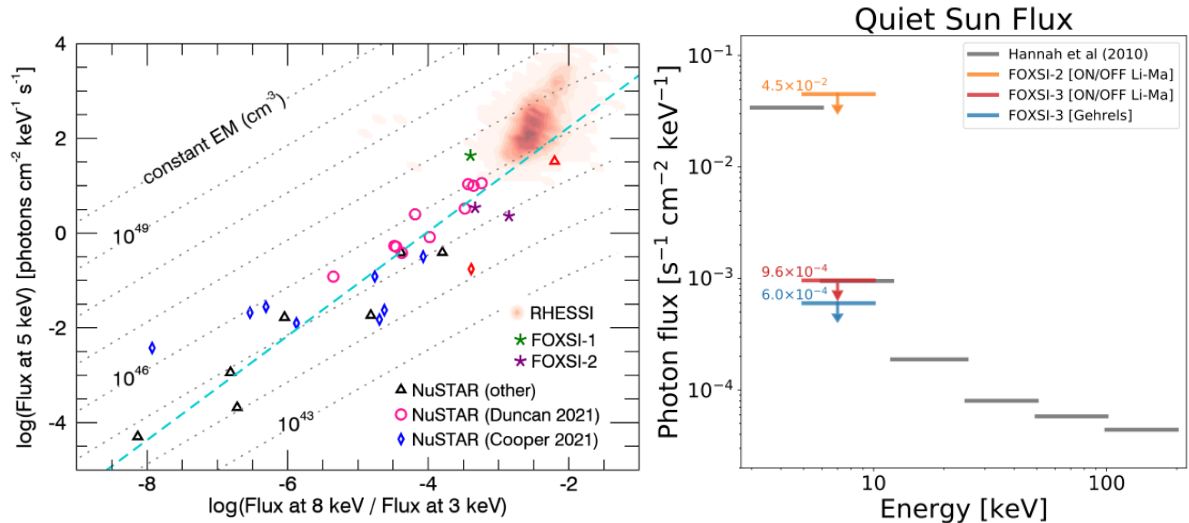

**Figure 4:** (Left) Brightness-hardness diagram from Duncan (2022). The red heat map in the upper right shows the microflares that *RHESSI* could measure, while individual data points show *NuSTAR* and *FOXSI* microflares. **This plot shows microflares measured down to 5 times fainter using focusing optics (*NuSTAR/FOXSI*) than indirect imaging (*RHESSI*)**. (Right) Upper limits on quiet Sun HXR flux from the *FOXSI-2* and *FOXSI-3* rocket experiments (colored lines), as compared with those from *RHESSI* (black lines). Using only 6 minutes of focused observation, similar or better limits are obtained as when using 12 days of *RHESSI* observing.

The thermal energy released from eruptive/heating events is often underestimated due to an isothermal approximation (Aschwanden et al 2015). It is challenging to precisely determine the amount of plasma at temperatures >5MK, which requires high sensitivity X-ray imaging and spectral observations. Multiple studies (e.g. Wright et al. 2017, Athiray et al 2020, Cooper et al 2021) have performed joint differential emission measure analysis of EUV and X-ray imaging/spectral data, demonstrating that adding HXRs can greatly improve the accuracy of high temperature plasma emission and thus provide better thermal energy estimates than with isothermal assumptions. The combination of focused HXR measurements with detailed *soft* X-ray measurements is especially powerful for this topic; see white paper by Oka et al. (2022).

Other white papers (Christe et al. 2022, Shih et al. 2022, Caspi et al. 2022) describe several different space mission implementations that include a direct HXR imager based on the *FOXSI* concept; please see those papers for the details of how such an instrument could be constructed right now. While *FOXSI* points the way to what can be accomplished with a directly focusing HXR solar instrument, NASA should not feel restricted to *solely* follow the path laid out by *FOXSI* technology. Several mirror fabrication methods are currently in development in the US and abroad, and these methods should be vetted to establish which one is best optimized for spectral response, angular resolution, cost, and mass. The *FOXSI* team has put in a concerted effort to make data, software, and tools publicly available, with data from all rocket flights hosted openly on the SDAC and with virtually all of its calibration and analysis software publicly available on GitHub.[1] The team has also published papers not only on the science but also the instrument development (e.g. Buitrago-Casas et al. 2020, Athiray et al. 2017, Ishikawa et al. 2016, Christe et al. 2016, Krucker et al. 2013), such that the larger community can benefit from the lessons learned during the *FOXSI* development.

---

[1] https://github.com/foxsi

## Summary

Studying electrons accelerated by solar flares is one of the most critical scientific paths for the next decade. These electrons contain an extraordinary amount of released magnetic energy; they influence how the entire rest of the flare develops; they link energy release sites in the low corona with nonthermal electrons that propagate throughout the heliosphere, and they additionally provide an important corollary to link heliospheric science with other astrophysical fields. *NuSTAR* and *FOXSI* have pointed the way to what a direct solar HXR imager could achieve, but most open questions cannot be solved with these platforms (in the case of *NuSTAR*, a non-optimized observatory observing only a few times per year, and in the case of *FOXSI*, a sounding rocket experiment observing for only a few minutes every few years). With over two decades since the last NASA solar HXR instrument and the technology already standing ready, the time is now right to implement a space-based mission optimized for and dedicated to the Sun.

| | Given name(s) | Family name(s) | ORCID iD | Email | Affiliation 1 | Affiliation 2 |
|---|---|---|---|---|---|---|
| Author 1 (no need to fill this column) | Lindsay | Glesener | 0000-0001-7092-2703 | glesener@umn.edu | University of Minnesota | |
| | Albert Y. | Shih | 0000-0001-6874-2594 | albert.y.shih@nasa.gov | NASA/GSFC | |
| | Amir | Caspi | 0000-0001-8702-8273 | amir@boulder.swri.edu | Southwest Research Institute | |
| | Ryan | Milligan | 0000-0001-5031-1892 | r.milligan@qub.ac.uk | Queen's University Belfast | |
| | Hugh | Hudson | 0000-0001-5685-1283 | hugh.hudson@glasgow.ac.uk | SSL UC Berkeley/ U of Glasgow | |
| | Mitsuo | Oka | 0000-0003-2191-1025 | moka@berkeley.edu | University of California Berkeley | |
| | Juan Camilo | Buitrago-Casas | 0000-0002-8203-4794 | juan@berkeley.edu | University of California Berkeley | |
| | Fan | Guo | 0000-0003-4315-3755 | guofan@lanl.gov | Los Alamos National Laboratory | |
| | Dan | Ryan | 0000-0001-8661-3825 | daniel.ryan@fhnw.ch | University of Applied Sciences and Arts N American University | |
| | Eduard | Kontar | 0000-0002-8078-0902 | eduard.kontar@glasgow.ac.uk | University of Glasgow | |
| | Astrid | Veronig | 0000-0003-2073-002X | astrid.veronig@uni-graz.at | University of Graz | |
| | Laura A. | Hayes | 0000-0002-6835-2390 | laura.hayes@esa.int | European Space Agency (ESA) | |
| | Andrew | Inglis | 0000-0003-0656-2437 | andrew.inglis@nasa.gov | NASA Goddard Space Flight Center | The Catholic University of Amer |
| | Leon | Golub | 0000-0001-9638-3082 | lgolub@cfa.harvard.edu | Center for Astrophysics | Harvard & Smithsonian | |
| | Nicole | Vilmer | 0000-0002-6872-3630 | nicole.vilmer@obspm.fr | LESIA-PSL/Paris Observatory | |
| | Dale | Gary | 0000-0003-2520-8396 | dgary@njit.edu | New Jersey Institute of Technology | |
| | Hamish | Reid | 0000-0002-6287-3494 | hamish.reid@ucl.ac.uk | University College London | |
| | Iain | Hannah | 0000-0003-1193-8603 | iain.hannah@glasgow.ac.uk | University of Glasgow | |
| | Graham S. | Kerr | 0000-0001-5316-914X | kerrg@cua.edu | Catholic University of America | NASA Goddard Space Flight Cen |
| | Katharine K. | Reeves | 0000-0002-6903-6832 | kreeves@cfa.harvard.edu | Center for Astrophysics | Harvard & Smithsonian | |
| | Joel | Allred | 0000-0003-4227-6809 | joel.c.allred@nasa.gov | NASA/GSFC | |
| | Silvina | Guidoni | 0000-0003-1439-4218 | guidoni@american.edu | American University | NASA GSFC |
| | Sijie | Yu | 0000-0003-2872-2614 | sijie.yu@njit.edu | New Jersey Institute of Technology | |
| | Steven | Christe | 0000-0001-6127-795X | steven.d.christe@nasa.gov | NASA/GSFC | |
| | Sophie | Musset | 0000-0002-0945-8996 | sophie.musset@esa.int | ESA ESTEC | |
| | Brian | Dennis | 0000-0001-8585-2349 | brian.r.dennis@nasa.gov | NASA/GSFC | |
| | Juan Carlos | Martinez Oliveros | 0000-0002-2587-1342 | oliveros@ssl.berkeley.edu | SSL/UC Berkeley | |
| | P. S. | Athiray | 0000-0002-4454-147X | athiray.panchap@uah.edu | University of Alabama in Huntsville | NASA MSFC |
| | Juliana | Vievering | 0000-0002-7407-6740 | Juliana.Vievering@jhuapl.edu | Johns Hopkins Applied Physics Laboratory | |
| | Stephen | White | 0000-0002-8574-8629 | stephen.white.24@us.af.mil | Air Force Research Laboratory | |
| | Amy | Winebarger | 0000-0002-5608-531X | amy.winebarger@nasa.gov | | |
| | James | Drake | 0000-0002-9150-1841 | drake@umd.edu | University of Maryland | |
| | Natasha | Jeffrey | 0000-0001-6583-1989 | natasha.jeffrey@northumbria.ac.uk | Northumbria University | |
| | Spiro | Antiochos | 0000-0003-0176-4312 | spiro.antiochos@umich.edu | University of Michigan | |
| | Jessie | Duncan | 0000-0002-6872-4406 | dunca369@umn.edu | NASA Goddard Space Flight Center | |
| | Yixian | Zhang | 0000-0001-8941-2017 | zhan6327@umn.edu | University of Minnesota | |
| | Meriem | Alaoui | 0000-0003-2932-3623 | alaoui@umd.edu | University of Maryland | |
| | Paulo J. A. | Simões | 0000-0002-4819-1884 | paulo@craam.mackenzie.br | Universidade Presbiteriana Mackenzie | University of Glasgow |
| | Marina | Battaglia | 0000-0003-1438-9099 | marina.battaglia@fhnw.ch | University of Applied Sciences and Arts Northwestern Switzerland | |
| | William | Setterberg | 0000-0003-2165-8314 | sette095@umn.edu | University of Minnesota | |
| | Reed | Masek | 0000-0003-2395-9524 | masek014@umn.edu | University of Minnesota | |
| | Thomas Y. | Chen | 0000-0002-0294-3614 | chen.thomas@columbia.edu | Columbia University | |
| | Marianne | Peterson | 0000-0002-9852-0869 | pet00184@umn.edu | University of Minnesota | |
| | Säm | Krucker | 0000-0002-2002-9180 | samuel.krucker@fhnw.ch | University of Applied Sciences and Arts N Univ. of California Berkeley | |
| | Manuela | Temmer | 0000-0003-4867-7558 | manuela.temmer@uni-graz.at | University of Graz | |
| | Pascal | Saint-Hilaire | 0000-0002-8283-4556 | shilaire@berkeley.edu | Univ. of California Berkeley | |
| | Vahe | Petrosian | 0000-0002-2670-8942 | vahep@stanford.edu | Stanford University | |
| | Trevor | Knuth | 0000-0002-4795-7059 | trevor.knuth@nasa.gov | NASA Goddard Space Flight Center | |
| | Christopher | Moore | 0000-0002-4103-6101 | christopher.s.moore@cfa.harvard.edu | Harvard-Smithsonian Center for Astrophysics | |